# 外化内不化，乘物以游心：在芯片体系结构的新黄金时代实现高性能计算程序的性能可移植

刘伟峰 武林平 徐小文 王昱人
中国工程物理研究院 北京应用物理与计算数学研究所

## 1) 引言

### 1.1 发展背景介绍

庄子在《知北游》和《人间世》中提出了一种人生价值观："外化而内不化，乘物以游心"。这是说为人处事，应该保持内心的庄重和秉持，又能平静温和的顺应自己所处的环境，只有这样才能够实现精神的自由和解放。庄子的这句话对于算法或程序开发人员来说同样有着重要意义。一个算法或程序如果在跨越多种系统时仍能自动适应系统架构，充分发挥系统的性能，实现性能的可移植，那么它们就更有可能获得广泛应用，创造出更大的价值[1]。高性能计算已经经历了数个阶段，作为高性能计算的一个重要目标，性能可移植的概念已经出现了很多年。早在1970年代，那时的程序员就已经开始面对将顺序程序从CDC-6600和CDC-7600计算机移植到矢量计算机Cray-1的挑战，而这最终促成了OpenMP编程标准[2]的出现。对于共享内存架构的计算节点来而言，OpenMP可以帮助程序在跨越多种操作系统、处理器架构以及编译器时获得良好的性能。接下来的90年代，程序员在经历了巨大的努力后完成了将程序从共享内存架构计算机向机群架构计算机的迁移，而这最终促成了MPI（消息传递接口）标准[3]的出现。进入新世纪后，超级计算机的节点逐渐开始大范围使用多核处理器，由于计算核心对共享内存的利用比传递消息效率更高，混合编程技术M+X （此处的M通常是指MPI，X可以为OpenMP[2]或者Cilk[4]等）变得重要。直到此时，程序开发人员仍然可以用同一种编程模型方便的实现程序在不同计算设备之间的移植。从以上角度看OpenMP以及MPI是性能移植的使能技术。综上，可以发现超级计算机体系结构的发展导致了编程模式的改变，编程模式必须不断进化以适应新的体系结构。

如果可以继续通过增加计算核心数量的方式提高处理器性能，那么简单、直观的M+X混合编程模型是实现性能移植的最佳选择。然而，随着Dennard缩放定律与摩尔定律的终结，当前通用处理器的发展遇到了严重的瓶颈，很难通过继续增加复杂的计算核心，以及提高计算核心频率来提升性能。一个解决方案是针对特定问题领域定制体系结构，为该领域提供显著的性能和能效收益，这也被称之为"领域特定架构"（DSA）。DSA的例子包括GPU、神经网络处理器、以及FPGA等。由于能更有效的利用特定领域的并行形式以及内存层次结构，与通用CPU相比，DSA可以实现更好的性能和更高的能效。以GPU为例，其在片上集成了众多流处理器核心，这些核心不包含乱序执行、cache控制等耗费资源的模块，

因此结构简单能耗较低。GPU借助SIMD并行技术,并通过快速的任务切换隐藏一组计算任务的访存延迟,实现了远高于通用处理器的计算吞吐量。GPU的特殊架构使其特别适用于大规模数据并行的计算任务。由于DSA只会加速某类程序,因此它们通常被称为加速器。在此背景下,将通用处理器与一个或多个协加速器在主板或片上相连,成为了增强单节点计算能力的普遍选择:通用处理器作为控制设备,负责计算任务的控制以及调度;协加速器则负责大规模并行计算任务或特定领域的计算任务。

表1. 2019年11月超级计算机TOP500榜单前10的系统

| 排名 | 系统/型号 | 计算节点结构 | 安装地点 | Linpack值/PFLOPS | 峰值/PFLOPS |
|---|---|---|---|---|---|
| 1 | Summit | IBM POWER9 22C 3.1GHz, NVIDIA Volta GV100 | ORNL | 148.6 | 187.66 |
| 2 | Sierra | IBM POWER9 22C 3.1GHz, NVIDIA Volta GV100 | LLNL | 94.6 | 125 |
| 3 | 神威太湖之光 | Sunway SW26010 260C 1.45GHz | 国家超算无锡中心 | 93.015 | 125.436 |
| 4 | Tianhe-2A | Intel Xeon E5-2692v2 12C 2.2GHz, TH Express-2, Matrix-2000 | 国家超算广州中心 | 61.445 | 100.0679 |
| 5 | Frontera | Intel Xeon Platinum 8280 28C 2.7GHz | 德克萨斯州高级计算中心 | 23.5 | 38.7 |
| 6 | Piz Daint | Intel Xeon E5-2690v3 12C 2.6GHz, NVIDIA Tesla P100 | 瑞士卢加诺国家超算中心 | 21.2 | 27.154 |
| 7 | Trinity | Intel Xeon E5-2698v3 16C 2.3GHz, Intel Xeon Phi 7250 68C 1.4GHz | LANL | 20.2 | 41.5 |
| 8 | ABCI | Intel Xeon Gold 6148 20C 2.4GHz, NVIDIA Tesla V100 SXM2 | 东京大学 | 19.88 | 32.56 |
| 9 | SuperMUC-NG | Intel Xeon Platinum 8174 24C 3.1GHz | 莱布尼兹计算中心 | 19.5 | 26.7 |
| 10 | Lassen | IBM POWER9 22C 3.1GHz, NVIDIA Tesla V100 | LLNL | 18.2 | 23.05 |

以表1中2019年11月超级计算机TOP500[5]榜单上排名前10的系统为例,可以发现这些超级计算机的计算节点普遍采用通用处理器加协处理器的异构架构,仅有第5和第9位的两台超级计算机采用了无协加速器的同构节点。具体到计算节点使用的通用处理器,主要有Intel Xeon、IBM Power、以及无锡江南计算技术研究所设计的申威;而协加速器则主要有Nvidia GPU、Intel Xeon Phi、以及中国国防科学技术大学的Matrix-2000。FPGA由于自身面临的挑战(相对稀少的片上资源以及较低的时钟频率),在当前的高性能计算领域应用并不广泛,同时由于超级计算机往往面向不同领域的程序,目前还没有Top10的系统安装专用的神经网络加速芯片。

## 1.2 在体系结构新黄金时代实现性能可移植的方式

由于当前同构超级计算机与异构超级计算机同时存在,且不同异构超级计算机采用的协处理器的架构也存在差异,如果开发人员在设计程序时没有将性能可移植纳入考虑,则当程序需要运行于采用其它架构的计算机上时,需要对其代码进行大量重写。因此,考虑

算法或程序的性能可移植，以便新的超级计算机在它们可用的那一刻就充分发挥其潜力，是当前高性能计算领域面临的主要挑战之一。程序性能可移植技术涵盖范围非常广泛，图1对当前几个重要的性能可移植技术的应用场景进行了总结:基于编程模型，编写满足性能可移植要求的代码；对于符合要求的串行代码，使用编译系统自动从代码中发现并行性，并面向不同的目标计算机生成相应代码；对于已经针对某种架构的计算机完成开发的代码，使用源码转换工具将其自动转换为可以在其它架构的计算机上运行的代码。

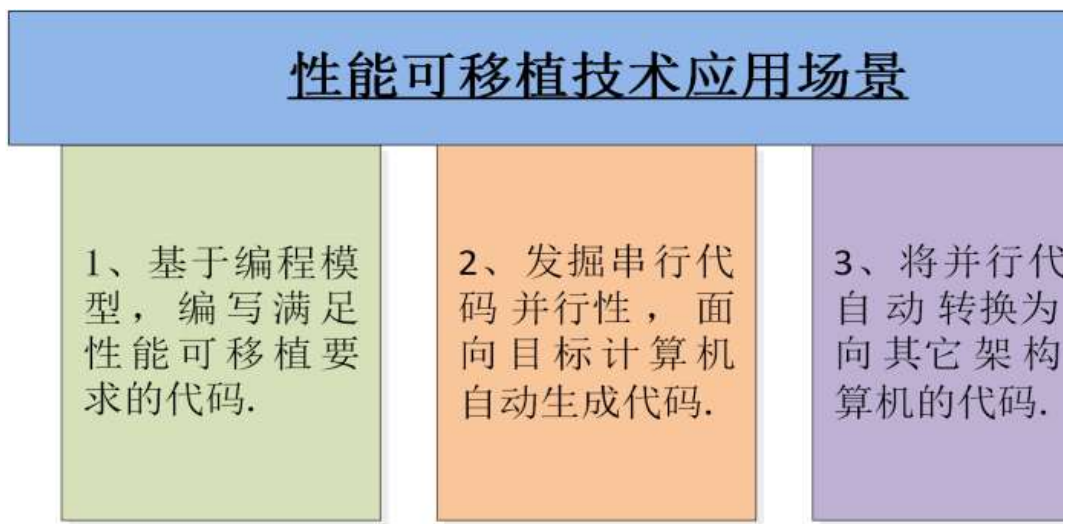

图1. 性能可移植技术的应用场景.

本文第二章将对满足性能可移植要求的模型进行重点介绍；第三章将对当前的串行代码自动并行化技术进行描述，并介绍一些使用这些技术进行代码移植的例子；第四章将总结当前的一些主流并行代码转换技术。本文最后还将介绍一些合理调用科学计算库函数，以提升程序性能的技术。

## 2、满足性能可移植要求的并行编程模型

在同构与异构超级计算机并存的情况下如何编写程序，是实现程序性能可移植首先要解决的问题。伴随着异构计算机的发展，原有的同构计算机并行编程模型(OpenMP，Cilk等)已不再适用，由此推动了大量与并行编程模型相关的研究。一个好的并行编程模型需能提供合理易用的计算设备抽象，使程序开发人员在不陷入复杂的硬件细节中的情况下，完成任务划分、任务映射、数据分布、计算同步、以及数据通信等任务；同时，编译优化系统作为并行编程模型的工具层，将开发人员编写的程序进行优化编译，生成可在目标计算设备上执行的文件。如表2所示，当前支持性能可移植的并行编程模型通常以编程接口、编译指导语句、以及编程语言的形式为程序开发人员提供计算设备抽象。本节将分别

对以上三类并行编程模型进行梳理总结。

表2. 当前主流并行编程模型

| 类型 | 编程模型 | 后端并行方式及支持的计算设备 | 支持性能可移植（否、是） | 索引 |
| --- | --- | --- | --- | --- |
| 编程接口 | OpenCL | CPU、Xeon Phi、GPU等 | 是 | [6] |
| 编程接口 | SkelCL | OpenCL | 是 | [7] |
| 编程接口 | Boost.compute | OpenCL | 是 | [8] |
| 编程接口 | Bolt | OpenCL | 是 | [9] |
| 编程接口 | SYCL | CPU、Xeon Phi、GPU等 | 是 | [10] |
| 编程接口 | DPC++ | CPU、Xeon Phi、GPU、FPGA等 | 是 | [11] |
| 编程接口 | C++ AMP | CPU、GPU等 | 是 | [11] |
| 编程接口 | Kokkos | OpenMP、CUDA、OpenCL | 是 | [12] |
| 编程接口 | RAJA | OpenMP、CUDA、OpenCL | 是 | [13] |
| 编程接口 | OP2 | MPI、OpenMP、CUDA、OpenCL | 是 | [14, 15] |
| 指导语句 | OpenACC | CPU、Xeon Phi、GPU等 | 是 | [16] |
| 指导语句 | OpenMP 4.X | CPU、Xeon Phi、GPU等 | 是 | [17] |
| 指导语句 | OpenHMPP | CPU、Xeon Phi、GPU等 | 是 | [18] |
| 指导语句 | Mint | GPU | 否 | [19] |
| 语言 | Chapel | CPU、GPU | 是 | [20, 21] |
| 语言 | X10 | CPU、GPU | 是 | [22] |
| 语言 | Halide | CPU、GPU等 | 是 | [44] |
| 语言 | TVM | CPU、GPU等 | 是 | [23] |
| 语言 | XLA | CPU、GPU等 | 是 | [24] |

## 2.1 基于编程接口的并行编程模型

为方便程序开发人员基于现有编程语言实现程序的性能可移植，当前有较多工作将对计算设备进行的操作以编程接口的形式提供给程序开发人员，从而方便开发人员基于现有语言使用各种不同计算设备。

## 2.1.1 标准编程接口

受支持不同网络类型的MPI消息传递接口启发，当前有许多组织以及机构尝试为多种架构的计算设备提供统一的编程接口，从而实现程序在不同计算设备之间的性能可移植。

伴随着NVIDIA GPU在超算领域的应用，NVIDIA于2007年推出了CUDA[25] 异构编程标准以帮助程序开发人员使用GPU对代码进行加速，然而CUDA仅能在以NVIDIA系列GPU 为协加速器的异构系统中使用，无法帮助程序在不同架构的计算设备之间实现性能可移植。为此，Khronos 组织于2008年提出了OpenCL[6]编程标准，尝试将各种处理器抽象为统一的模型，Intel的Xeon Phi、NVIDIA GPU、AMD GPU、以及神威太湖之光超级计算机使用的SW2601处理器都对OpenCL进行了支持[26-29]。OpenCL 的编程接口较为底层，造成程序开发效率降低。为了提高开发效率，SkelCL[7]、Bolt[9]、以及Boost.compute[8]等以模板库的方式封装并简化了OpenCL的数据与任务管理操作，其中Bolt[9]以及Boost.compute[8]还基于OpenCL实现了一些常用的并行算法供编程人员使用。Khronos 组织于2014年提出了 SYCL [10]编程标准， 作为OpenCL之上的抽象层，SYCL 标准支持高维数组，简化了OpenCL中的数据与任务管理操作。DPC++[11]是Intel基于SYCL设计的编程标准，其对SYCL进行了适当扩展，支持跨CPU和协加速器的数据并行编程，同时还支持面向特定的协加速器进行调优。C++ AMP[30]是微软提出的编程标准，其比OpenCL拥有更高层次的计算设备抽象，因此代码非常简洁，只需要在for循环中指定计算域以及内嵌的lambda函数就能完成并行计算。C++ AMP的支持多维数组以及内存数据布局（Row/column major, AOS, SOA），依托于Visual Studio的支持，以及微软为C++ AMP专门开发的并行算法库，其为Windows开发者提供了良好的支持。

## 2.1.2 适配不同标准编程接口的模板函数

每一种并行编程标准都有对其进程支持的计算设备，为了能够使用尽可能多的计算设备，当前有一些编程模型通过用统一的接口适配不同的并行编程标准，达到使程序在不同架构计算设备之间性能可移植的目标。

Kokkos[12]是一套帮助程序实现在不同架构的计算设备之间性能可移植的C++模板库，其核心是数组的抽象容器view以及数据并行操作的抽象类functor。通过对不同的设备内存进行抽象，view不仅支持多维数组，还把内存对齐、下标映射、内存数据布局、以及访问控制等封装起来。functor类包含程序员定义的数据操作函数以及这些函数需要操作的数组容器，其主要作用是将设备的执行空间与存储空间连接起来。Kokkos目前支持的编程标准包括OpenMP，Pthreads、CUDA、以及OpenCL等。当前著名的并行计算基础库Trilinos[31]，其中并行线性代数容器Tpetra的底层就是基于Kokkos实现的。

RAJA[13]同样是一套帮助程序实现在不同架构计算设备间性能可移植的C++模板库，其参考了劳伦斯利物莫国家实验室的一系列结构以及非结构网格程序的特点，认为这些程序中对网格进行遍历操作的for循环的各循环实例之间不存在数据依赖，可以并行执行。RAJA将这些for循环抽象成为模板函数，模板函数的参数包括用户指定的循环体执行策略以及目标编程标准等。RAJA还引入了Indexsets的概念，可以自动将循环空间划分成为不同的分段，最终在编译的过程中会为每一类分段分别生成代码。RAJA目前支持的编程标准包括OpenMP、Pthreads、以及CUDA等。

OP2[14, 15]是专门针对非结构网格程序设计的编程模板，其前身是Oplus[32, 33]。OP2将非结构网格程序划分为4部分：集合（边的集合，网格的集合等）、集合相关的数据、集合之间的映射关系（边与网格的对应关系）、以及对集合的操作。其认为所有非结构网格程序都是对相关集合进行算数操作，体现在代码上就是使用for循环对集合的元素进行遍历，直接或通过映射关系间接访问并操作集合的数据。使用OP2可以实现程序的节点间以及节点内并行，其中节点间并行基于MPI编程标准实现，节点内并行根据后端计算设备可以基于Pthread、OpenMP、CUDA、OpenCL等编程标准实现。在实现节点间并行时，OP2框架使用ParMETISs[34]以及PTScotch[35]对整个计算域的网格进行划分以确保节点之间的负载平衡；在实现节点内并行时，OP2采用图着色算法实现节点内处理单元的任务分配及负载平衡。OP2框架还可以根据不同后端计算设备的特点，选择不同的数据布局方式（AOS或SOA）。早期的OP2框架基于ROSE源码编译器实现源码编译，而最新的OP2框架[36]采用LLVM-Clang重新实现了源码编译功能。

## 2.2基于编译指导语句的编程模型

对于数量庞大的遗产代码，使用插入编译指导语句的方式可以帮助开发人员使用不同计算设备对其进行加速。OpenACC[16]是由Cray、PGI、以及英伟达发起的一个编程标准，与OpenMP类似，其允许将编译指导语句插入Fortran,C和C++程序的代码中，以帮助编译器将计算任务调度到协加速器上。目前对OpenACC进行支持的设备包括NVIDIA GPU, AMD GPU, SW26010, Intel Xeon Phi,FPGA等。OpenACC的出现极大的方便了程序开发人员使用协处理器加速数量庞大的遗产代码。与OpenACC类似，HMPP[18]是由CAPS组织发起的一种基于编译指导语句的编程标准，其支持C和Fortran两种语言。目前NIVIDA 系列GPU对HMPP进行了支持。HMPP编译器可以根据#pragma编译指导语句生成在相应计算设备上执行的二进制文件。自从OpenMP 4.X[17]开始，OpenMP中也引入了用于协处理器加速的指导语句，目前NVIDIA GPU, AMD GPU, Intel Xeon Phi、以及IBM Power处理器都对OpenMP 4.X进行了支持。OpenMP对底层的抽象比OpenACC要复杂，导致其使用难度要高于OpenACC。Mint[19]是为没有CUDA编程经验的人员设计的编程模型， 其可以帮助他们使用

简单的5条编译指导语句将科学计算中广泛存在的串行Stencil代码转换为CUDA代码。Mint基于ROSE源码编译器实现了以上源码转换，在转换的过程中还集成了访存优化策略，目前Mint只能生成CUDA代码，但其经过简单扩展，就可以实现对OpenCL的支持。Niklas在他的工作中使用LLVM前端Clang重新实现了Mint[37]。

## 2.3 基于语言的编程模型

当前还有许多工作设计新的编程语言，这些编程语言对并行编程所需的常用操作进行总结，从而简化并行程序的开发难度。过去的很长一段时间，程序中描述算法的逻辑被性能优化策略掩盖。针对特定计算设备进行性能优化后的代码基本被固定，不再具有良好的可维护性，代码在进行跨计算设备移植时需要大范围重写。因此，程序开发人员往往采用保守的方案，把性能优化留到最后一步。为了提高代码的性能可移植性，有研究将算法描述和代码生成解耦，使用简单的语法对算法进行描述，然后再针对具体的计算设备生成相应的优化代码，这样在不同的计算设备间移植算法时只需要对算法的代码生成模块进行替换。本节将分别介绍算法描述与优化策略混合以及解耦的编程语言。

### 2.3.1 算法实现与优化混合的编程语言

过去一段时间，有一些面向大规模并行计算的程序语言被设计出来，经过适当的扩展，这些编程语言目前也对协加速器提供支持。这些编程语言可以用于实现任何算法，但由于这些算法往往较为复杂，因此这些程序设计语言较难做到将算法描述以及优化完全解耦。

X10[22]是IBM在美国DARPA的HPCS（High Productivity Computing Systems）项目中提出的一种编程语言，其基于JAVA进行扩展，使用异步划分地址空间模型(APGAS Model)管理CPU以及GPU，删除了JAVA中的并行控制部分，引入了新的并发控制库。X10通过简单扩展后使用自身的语义替代CUDA的相关语义，简化了GPU编程[38]。Chapel[21, 39]是Cray在HPCS项目中提出的一种并行编程语言，其创造性的提出了全局视角的并行编程模式、数据定义与算法实现分离、以及在不同计算设备之间实现性能可移植。为了实现相同代码在不同计算设备上执行，Chapel提供了一系列供预定义的接口[40, 41]，通过实现这些接口，程序开发人员可以指明目标计算设备的数据管理方式以及循环中各循环实例的并行方式。通过以上方式，Chapel程序目前可以使用多核CPU以及NVIDIA GPU[42]进行执行。

### 2.3.2 算法实现与优化解耦的编程语言

目前，在高性能计算的某些特殊领域，已经有某些专用语言可以将程序的算法描述和

代码生成完全解耦。该类领域专用语言能够帮助程序开发人员对领域算法进行描述,并在不关心底层细节的情况下,仅通过调用优化接口就完成算法优化以及代码生成。

信号处理领域的SPIRAL框架[43],实现了算法描述与代码生成优化分离,为信号处理算法提供了从顶层算法描述到底层代码实现的映射。具体的,SPIRAL框架可以划分为算法描述层、代码生成层、以及测试优化层3层。其算法描述层使用信号处理语言SPL对信号处理算法进行描述。代码生成层负责将SPL语言描述的算法转化为C或Fortran代码,在代码转化的过程中会涉及大量优化参数的选择,如循环展开层数以及循环分块大小等,优化空间非常巨大,例如使用SPL描述的DCT-264算法就存在高达种转化选择。如果使用穷举的方法从优化空间中选择最优的代码转化方案,则时间开销是不可接受的。为此,SPIRAL框架的测试优化层采用了自动优化策略,基于进化算法或机器学习算法控制代码生成层生成代码。SPIRAL框架目前可以使用多种不同架构的多核处理器,暂不支持GPU等加速设备。

图像处理领域的Halide[44]语言,实现了图像处理算法的描述与优化的分离。一个Halide程序由算法描述以及调度两部分组成,算法描述部分可以高效实现常见的图像处理算法,如去除红眼、柔化阴影或增加饱和度等,调度优化部分则根据目标计算设备的架构指定一些优化策略,如计算分块、并行方式、向量化方式等。使用Halide语言的程序开发人员如果想要将程序移植到采用不同架构的计算设备上,只需对调度优化部分进行变更。不同的计算设备有着不同的优化方法,Halide认为所有的优化方法归根结底都是对存储或计算顺序的控制,因此其提供了统一的调度优化接口,控制算法执行过程中访存和计算顺序,帮助程序开发人员描述性能优化方案。一旦调度优化策略确定,Halide将自动结合算法描述、调度优化策略、以及目标计算设备,生成相应的代码。Halide目前可以使用的计算设备包括CPU、GPU、DSP,FPGA、以及ASIC。Google的Pixel 2手机项目使用Halide实现了HDR+算法,使其运行于IPU芯片上。使用Halide提供的调度优化接口,程序员可以方便的尝试不同的优化策略,并发现一些有效的优化策略,但是由于需要人工不断尝试,因此效率并不高。为了提高优化效率,Hailde借鉴PolyMage工具,实现了自动优化策略。

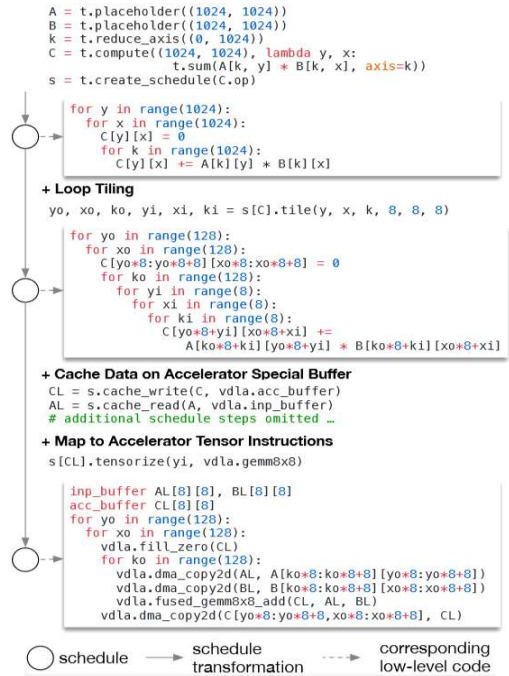

图1：使用TVM解耦矩阵乘法算法描述以及算法优化[23]

目前深度学习领域的研究十分活跃，每天都有新的神经网络算子（层）被提出来，以期望进一步提升模型的精确度。同时，由于越来越多的厂商开始设计制作神经网络芯片，运行神经网络模型时会有越来越多的后端计算设备可供选择。由于既要支持层出不穷的神经网络算子（层），又要保证新的神经网络算子（层）可以运行于不同的后端计算设备之上，深度学习框架面临着极大的挑战，由此导致了众多深度学习编译器项目的出现。TVM[23]是一套完整的深度学习优化框架，其包括神经网络模型的图优化层以及算子优化层。图优化层会执行算子合并等操作以对计算图进行优化；算子优化层会针对特定后端计算设备，为计算图中的算子生成优化的代码。算子优化层借鉴了Halide语言的思想，将算子描述和代码生成解耦，使用一种专门设计的张量语言描述算子，同时从不同的后端计算设备抽象出了一整套优化调度优化接口。图1描述了使用张量语言对矩阵乘算法进行描述，以及通过调度优化接口对算法进行优化。当确定了后端计算设备之后，算子优化层采用自动优化的策略，不断在用户指定的搜索空间中搜索调度优化策略。TVM目前可以使用的计算设备包括CPU、GPU、DSP，FPGA等。

与TVM类似的还有Google开发的XLA深度学习编译器，XLA的优化流程可以分成目标无关优化和目标相关优化。其目标无关优化层对HLO描述的Tensorflow计算图进行优化；而目标相关优化层则将Tensorflow的算子编译到不同的计算设备上（Nvidia GPU，张量处理器单元TPU等）。

## 2.4 编程模型小结

当前异构并行编程模型通常以编程接口、编译指导语句、以及编程语言的形式为程序开发人员提供计算设备抽象，使程序能够充分利用各种计算设备的资源。

在编程接口方面，OpenCL作为较早被提出的标准编程接口，已经得到了较为广泛的应用。由于OpenCL编程接口较为底层，为了提高程序开发效率，一些工作将OpenCL与底层细节相关的操作进行了简化，并把一些常用的算法用OpenCL进行了实现。Kokkos、RAJA、以及OP2则可以使用统一的编程接口适配不同的并行编程标准，从而帮助程序开发人员使用更多的计算设备。

在编译指导语句方面，OpenACC、HMPP、OpenMP 4.X可以帮助开发人员快速使用不同协加速器对代码进行加速。Mint则可以帮助程序开发人员使用简单的5条编译指导语句将串行Stencil代码转换为CUDA以及OpenCL代码，从而实现代码在不同计算设备间的移植。然而，由于编译指导语句能够提供给编译器的信息不够丰富，因此其代码加速效果往往不如底层编程接口好。

在编程语言方面，Chapel以及X10等并行计算语言经过简单扩展，就可以使用多种不同的计算设备。同时在某些特定领域，领域专用语言已经可以将程序的算法描述和代码生成解耦，该类语言能够帮助程序开发人员在不关心底层细节的情况下，通过调用优化接口完成算法优化以及代码生成，较好的平衡了编程效率以及优化效果。

# 3、 串行代码自动并行化

一直以来，并行计算领域的圣杯是程序开发人员只需要编写完全串行的代码，由编译系统或运行时环境自动从代码中发现并行性，并面向目标计算设备生成相应代码，通过串行代码自动并行化直接实现代码的性能移植。

## 3.1 基于多面体编译技术的串行代码自动并行化

对于某些符合要求的代码，多面体编译技术已经可以将其自动并行化。多面体编译技术中的多面体指的是在循环边界的线性约束条件下，各循环实例被包围在一个空间凸多面体内，在不违反依赖关系的情况下，对各循环实例的空间位置进行采用相同的调度优化策略（仿射变换）后，各个循环实例仍然被被包围在一个空间凸多面体内。

多面体编译工具主要由抽象分析、调度优化、以及代码生成三个部分组成。抽象分析部分作为多面体编译工具的前端，其首先从输入的循环代码中识别迭代空间和访存映射；然后，根据迭代空间和访存映射，计算循环实例之间的依赖关系。调度变换作为多面体编译工具的中间优化层，其通过调用线性整数规划函数，在满足循环实例依赖关系的前提下，结合后端计算设备的架构特点，求一个充分挖掘代码并行性以及数据局部性的优化策略（通

过仿射变换改变循环实例的执行顺序）。代码生成部分作为多面体编译工具的后端，其以迭代空间、依赖关系、以及调度优化策略作为输入，生成抽象语法树并最终将抽象语法树转变成代码。

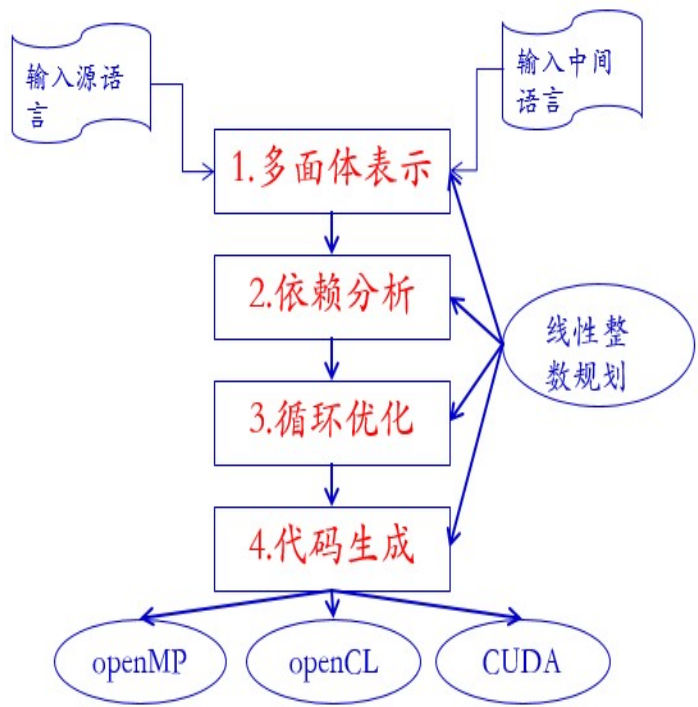

图2：基于多面体编译技术的一般编译流程[45]

当前的主流编译器已逐渐集成了多面体编译技术，如 GCC 集成了Graphite 框架[46, 47]，LLVM 集成了Polly框架[48]，从而针对特定计算设备改进编译结果的并行性以及访存性能。多面体编译技术不仅可以作为一个编译模块嵌入到通用编译器中,还可以作为独立的编译工具， 实现从源程序到源程序的翻译流程。面向共享内存的多核架构，多面体编译工具Pluto[49]可以将串行循环代码自动转换成为OpenMP代码。面向GPU等计算设备，Konstantinidis等人开发的多面体编译工具[50]可以将串行代码转换成CUDA代码，该工作后被PPCG[51]超越，PPCG能够支持Pencil语言[52],基于PPCG+Pencil的组合，程序开发人员可以采用"编译制导+多面体编译"的模式对程序进行优化。此外，Polly-ACC[53]基于Polly框架实现了从中间语言到中间语言的翻译,并最终可以生成OpenCL或CUDA目标代码。由于Polly-ACC的操作对象为源代码编译后的IR中间语言，其可以对多种语言编写的源代码提供优化。

由于多面体编译技术可以自动发掘程序并行性和数据局部性，极大降低了程序开发人员使用不同架构计算设备的门槛，其开始得到了越来越广泛的关注。但多面体编译技术并不是机器猫的神奇口袋，其并不能自动为所有的循环代码找出并行优化方案。首先，多面

体编译技术严格要求输入程序的循环边界为各循环变量的线性组合；其次，多面体编译技术对代码的控制流程有较高要求，对复杂条件判断语句的支持不足；最后，多面体编译技术的依赖分析部分要求数组下标需要为循环变量的线性组合，对复杂下标的支持不足。由于以上诸多限制，多面体编译技术目前常被用于对符合要求的Stencil代码进行优化。

由于目前深度学习领域中的神经网络算子（层）往往可以用符合多面体编译技术要求的Stencil代码描述，多面体编译技术已经开始被用于深度学习领域的算子优化。TVM以及XLA的算子优化层对于有体系结构背景知识的程序开发人员来说是一种高效的方案，但对于大多数深度学习算法设计人员来说却有着一定的门槛。针对以上问题，Facebook开发了Tensor Comprehensions（TC）[54]，以帮助深度学习算法设计人员更方便的将代码部署到不同的后端计算设备上。与TVM和XLA一样，TC同样定义了一种张量表达语言，以方便算法设计人员对算子的计算进行描述，与TVM和XLA不同的是，TC把算法优化部分用多面体编译技术实现了，用户不需要编写任何与优化相关的语句。以GPU为例，TC可以基于多面体编译技术实现循环的融合、分裂、分块、以及自动并行，同时还确保数据在复杂存储层次中正确移动。为了寻找最优的调度优化策略，TC采用进化搜索算法对海量方案进行评估，并从中选择性能最佳的方案。目前TC能够使用的计算设备主要为众核CPU以及GPU。

## 3.2基于遗传编程的串行代码自动并行化

Koza于1992年提出了遗传编程[55, 56]，遗传编程属于进化算法，其继承了遗传算法的基本思想，即从基数较为庞大的原始、粗糙的程序种群中通过评估适应性选择父系，然后进行变异、交叉等遗传操作生成新一代程序种群，再判断终止条件决定是否需要继续迭代、生成下一代种群。Koza在论文中使用语法树表示程序代码，程序代码的变异操作通过改变语法树节点上的操作、增加或删除分支、或用一棵全新的树来替换其子树完成；而程序代码的交叉可以通过用一棵语法树的分支取代另一棵语法树的分支完成。遗传编程在生成复杂控制代码方面取得了较大的成功。文法进化算法[57]是基于文法的遗传编程，其通过预先构建的编程语言的BNF范式，将程序代码的基因型（二进制串）与表现型（语法树）对应了起来，通过简单替换语言的BNF范式，就可以用相同的算法生成不同语言的代码。图3对使用BNF范式将程序代码基因型与表现型进行对应的过程做了描述。Langdon等人在他们的工作中[58]使用文法进化算法，将gzip程序中被频繁调用的串行函数longest_match转换成CUDA代码。该工作基于CUDA提供的示例程序scan_naive_kernel.cu构造BNF范式，并以构造的BNF范式为基础不断进化出新的CUDA代码。进化生成的代码被使用nvcc编译器编译并链接到gzip，然后使用测试数据验证生成的代码的正确性。使用该方法，最终进化出了能通过所有测试用例的CUDA代码，该CUDA代码后期在上百万次运行过程中全部都能产生正确结果。使用同样的方法，Gopinath等人在他们的工作中[59-62]将串

行代码自动转换成运行于多核计算设备的OpenMP代码，同时他们还将代码进化的过程并行化，以加快代码生成速度。使用遗传编程面临的最主要问题是最终进化出的代码的可解释性较差，因此无法确保代码在所有可能的情况下都能输出正确的结果。

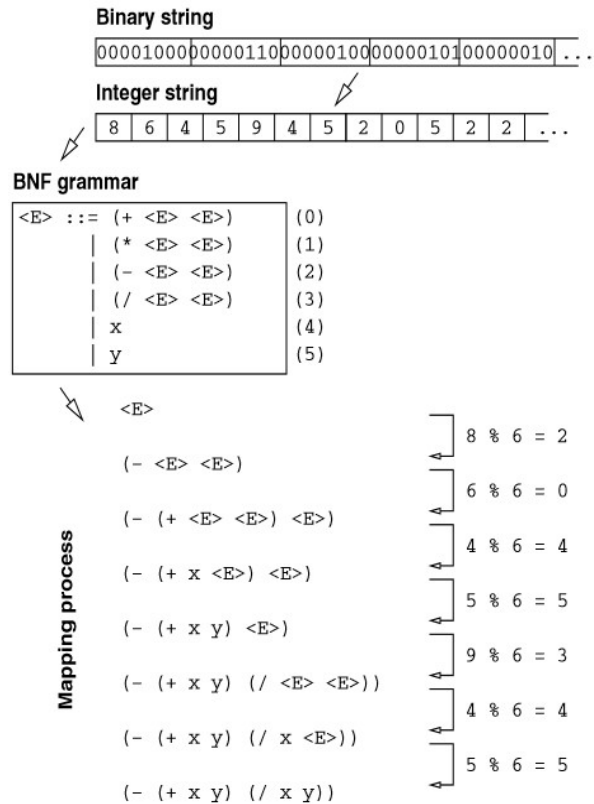

图3：文法进化算法使用BNF范式将程序代码的基因型与表现型对应的过程。基因型二进制串被转换成整型串，每个整型对应基因型二进制串中的8个bit。算法依次使用整型串中的数字选择BNF的生成规则，并对起始符号不断应用这些选中的规则[63]

## 3.3基于编译信息的串行代码自动并行化

Nawata等人开发的APTCC[64]采用源码分析技术，将可并行的c语言for循环语句转换成CUDA代码。在执行以上转换的过程中APTCC分析需要从主机端传送到设备端的数据，并自动将for循环转换成在GPU流处理器上执行的函数。APTCC无法自动判断for循环语句是否可并行，因此需要使用者指明目标for循环。解放军信息工程大学的赵荣彩团队借助open64编译器提供的一些编译信息，自动对某些符合并行条件的for循环添加OpenACC编译指导语句，从而完成代码移植工作[65]。由于open64仅能发现当前形式下for循环的可并行性，并不能对代码做更进一步的变换以及调度，从而导致该框架不能进一步优化代码的并行性以及数据本地性。

## 3.4 串行代码自动并行化技术小结

串行代码自动并行化是实现代码的性能可移植最直接的方法，然而由于技术的限制，当前的串行代码自动并行化面临一些困难。首先，多面体编译技术由于严格要求输入程序的循环边界以及数组下标为各循环变量的线性组合，且对复杂条件判断语句的支持不足，目前常被用于Stencil代码优化；其次，使用遗传编程将串行代码并行化，本质上是用类似于机器学习的方法不断生成代码对测试数据进行拟合，因此无法确保最终生成的代码在接收测试用例之外的数据后也能输出正确的结果；最后，当前编译器能够提取的代码的优化信息往往有限，因此基于编译信息的串行代码自动并行化技术不能像多面体编译技术一样通过改进代码，进一步发掘代码的并行性以及数据本地性。

# 4、面向性能可移植的并行代码自动转换工具

对于已经针对某种架构的计算设备完成并行化的代码，可以使用代码转换工具将其自动转换为可以在其它架构的计算设备上运行的代码，从而实现代码在不同架构的计算设备之间的移植。由于已经通过特定方式指明了代码的并行策略以及访存方式，因此无需对代码进行复杂的分析以确定代码的数据依赖关系以及可并行性，只需要结合目标计算设备的架构特点，重新实现当前代码中已有的并行策略以及存储优化方式。

## 4.1 同构到异构架构的自动代码转换技术

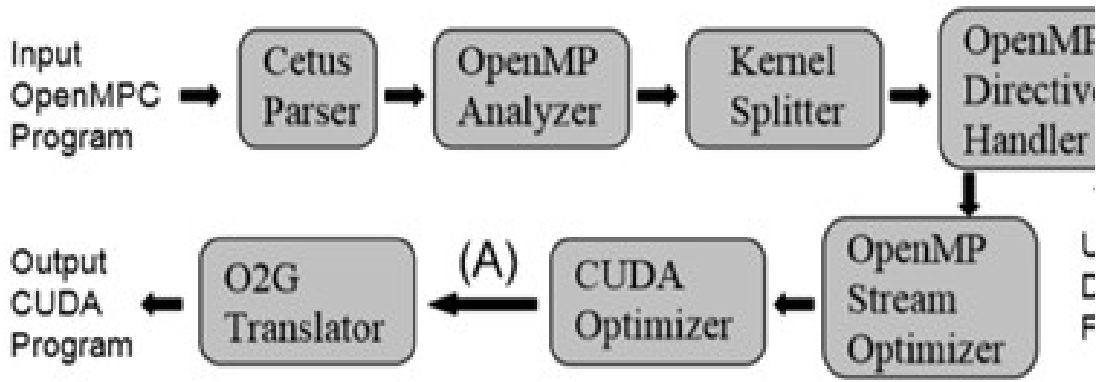

图4：OpenMPC源码转换以及优化流程[66]

OpenMP被广泛用于共享内存计算机的并行编程，OpenMP可以高效表达for循环各循环实例之间的并行，其fork-join模型中主线程以及工作线程的概念也能与CUDA编程模型中运行于CPU的主线程以及运行于GPU的工作线程较好的对应起来。Seyong等人开发的源码编译器Cetus[67]，可以自动将OpenMP代码转变成CUDA代码，从而实现程序在不同架构计算设备之间的移植。在执行代码转换的过程中，Cetus首先识别代码中的OpenMP指导语句，

并将其转换成对应的CUDA编程模型中的语句；然后，Cetus从程序中抽取出核心代码，并将其转换为CUDA的核函数；最后，Cetus解析出GPU要处理的数据，并插入CUDA数据传输语句。Cetus针对CPU以及GPU的架构特点，采用了新型工作负载分配策略、循环交换、循环合并、线程私有数组按列主序扩展、以及数据缓存等一系列技术优化GPU线程的访存。Cetus基于Java语言实现，其通过构造抽象语法树并对其进行一系列操作完成源码转换。由于Cetus的成功，Seyong等人提出了OpenMPC[66]，OpenMPC在Cetus的基础上定义了几类标记，分别用于让用户指定一些优化细则、设置优化参数、以及为环境变量赋值，以供其自动调优框架进行调优。图4描述了OpenMPC源码转换以及优化流程。Cetus以及OpenMPC的源码转换机制复杂，同时由于生成的是CUDA代码，因此生成的代码能使用的计算设备类型有限。王燕燕等人在他们的工作中[68]，使用Clang编译器完成了OpenMP代码到OpenCL代码的自动转换，在源码转换的过程中还采用了部分与Cetus相似的访存优化技术。该工作使用了Clang编译器以简化源码转换流程，同时由于生成的是OpenCL代码，因此能使用各类不同架构的计算设备。

## 4.2 异构到同构架构的自动代码转换技术

与以上工作完全相反，并行代码自动转换工具MCUDA[69]可以将CUDA代码自动转换成在共享内存架构计算机上运行的代码。GPU中不同线程块之间完全独立，因此MCUDA将同一个线程块中的线程全部映射到同一个处理器核心，以避免计算核心之间的同步操作。对每一个GPU线程块，MCUDA会构造一个for循环顺序执行块内所有GPU线程的代码，如果GPU线程块代码中有同步点，在构造for循环时会删除同步语句，并使用循环分裂技术从同步点处将for循环拆分。MCUDA在构造for循环时，会对GPU线程块中每个GPU线程的私有数据的作用范围进行分析，对于作用范围超出for循环的变量，MCUDA会将其扩展成为数组，每一个循环实例对应一个数组元素，而对于作用范围仅限于for循环的变量则不扩展。在进行代码转换的过程中，MCUDA将一个GPU线程块要访问的数据存储在计算核心独有的L1级缓存中，而对于各线程块同时访问的数据则存储在各计算核心共享的L2级缓存中。

Joel等人开发了Clacc[70]，可以将OpenACC代码自动转换成OpenMP代码，使用LLVM编译器对转换后的代码进行编译,可以实现LLVM编译器对OpenACC的支持。相比于OpenMP,OpenACC对底层硬件进行了更高程度的抽象，因此该转换过程与编译器的自上而下的编译流程相似。首先，Clacc使用LLVM的前端编译器Clang创建包含OpenACC节点的虚拟语法树，然后，使用专门开发的acc2omp工具将OpenACC的虚拟语法树转换成OpenMP的虚拟语法树，并将该虚拟语法树转换成LLVM的中间语言，最后，使用LLVM后端生成可执行代码。目前，Clacc将OpenACC代码转换为OpenMP代码的过程中，没有引入过多的优化策略。与Clacc类似的还有Nawrin Sultana等人的工作[71]，他们将设计的替换算法集成到

Eclipse C/C++ Development Tools中，实现了将运行于GPU的OpenACC代码转换为OpenMP代码。

### 4.3 并行代码自动转换工具小结

并行代码自动转换可将数量庞大的遗产并行代码迅速转换为可以在不同架构的计算设备上运行的代码，因此其有着重要的研究价值。由于只需要结合目标计算设备的架构特点重新实现并行代码中已有的并行策略以及存储优化方法，该研究方向不存在理论上的难题。同时，随着Clang等源码分析工具的不断发展，开发并行代码转换工具的技术门槛也在不断降低。综上，该研究方向将成为性能可移植领域的热点。

# 5、其它

一个对各种不同架构计算设备都提供良好支持的科学计算库可以帮助开发人员节省大量时间和精力，这是实现程序性能可移植，并提高生产效率的有效方式。在过去几十年中，出现了很多应用广泛且针对不同架构计算设备进行优化的高性能计算函数库，如基本线性代数例程（BLAS）、高扩展线性代数库（scalable LAPACK）、以及西方最快傅里叶变换（FFTW）等。尽管当前许多程序广泛使用了这些针对特定计算设备高度优化的科学计算库，但是仍然需要大量手写代码对一个库函数的输出进行预处理（数据布局更新，存储分配），从而将其转换为另外一个库函数的输入。胶水代码会影响程序的可移植性，这是因为胶水代码往往是针对特定计算设备编写的，因此在将其移植到新的计算设备上时，也需要针对新的计算设备的架构进行优化。另外在使用协加速器的计算节点上，胶水代码还会导致不必要的主机内存与设备内存之间的数据移动。图5(a)中基于fftwf_plan_dft_1d函数实现快速傅里叶变换，代码中存在较多胶水代码以循环调用fftwf_plan_dft_1d函数并进行数据重排，而图5(b)中使用fftwf_plan_guru_dft函数对图5(a)中的代码进行重写，胶水代码的数量大大减少。实验结果表明图5(b)中的代码相对于图5(a)中的代码有着显著的性能提升[72]。

```
1.   //planning
2.   fftw_plan plan_fft = fftw_plan_dft_1d(N_DOP,
3.       &datacube_pulse_major_padded[0][0][0],
4.         &datacube_pulse_major_padded[0][0][0],
5.         FFTW_FORWARD,
6.         FFTW_PATIENT);
7.
8.   // multiple 1-D FFTs
9.   for (chan = 0; chan < N_CHAN; ++chan)
10.     for (range = 0; range < N_RANGE; ++range)
11.       fftw_execute_dft(plan_fft,
12.         &datacube_pulse_major_padded[chan][range][0],
13.         &datacube_pulse_major_padded[chan][range][0]);
14.
15.   // data layout transform
16.   for (chan = 0; chan < N_CHAN; ++chan)
17.     for (range = 0; range < N_RANGE; ++range)
18.       for (dop = 0; dop < N_DOP; ++dop)
19.         doppler_data_cube[chan][dop][range] =
20.           datacube_pulse_major_padded([chan][range][dop]);
```

```
1.   //planning
2.   fftwf_plan plan_fft;
3.   const fftwf_iodim dims[1] = {{N_DOP, 1, 1}};
4.   const fftwf_iodim howmany_dims[2] =
5.       {{N_RANGE, N_DOP, N_DOP},
6.        {N_CHAN, N_RANGE * N_DOP, N_RANGE * N_DOP}};
7.
8.   // FFT operation
9.   plan_fft = fftwf_plan_guru_dft(1, dims,
10.      2, howmany_dims,
11.      datacube_pulse_major_padded,
12.      doppler_data_cube,
13.      FFTW_FORWARD, FFTW_PATIENT);
14.
15.  // execute plan
16.  fftw_execute(plan_fft);
```

（a）快速傅里叶变换原始代码　　　（b）代码融合后的傅里叶变换代码

图5：快速傅里叶变换代码优化[72]

以上示例显示了合理使用科学计算库函数，可以显著提升程序的性能。为了帮助程序开发人员选择合适的科学计算库函数进行计算，李韦等人在他们的工作中开发了一个源码编译器[72]以优化科学计算库函数的使用。首先，他们对The Spiral Language[73]进行改进，设计了一套通用的领域语言，该领域语言支持基本的数学运算功能，还包含了一些高性能计算库函数源语，这些源语与FFTW以及BLAS等常用高性能计算库中的函数一一对应，同时该领域语言定义了一些符号用于抽象程序的数学计算行为。然后，他们基于GAP4[74]开发了编译器前端，该前端能够将图5（a）中的代码编译成图6(a)中所示使用领域语言描述的代码，图6(a)中的代码按照优化规则优化后，被转换为图6(b)中的代码。最后，他们开发了编译器后端，该后端能将图6(b)中的代码转换为图5（b）中的代码。

$$(I_{N\_CHAN} \otimes L_{N\_RANGE}^{N\_RANGE \cdot N\_DOP}) \circ$$
$$(I_{N\_CHAN} \otimes (I_{N\_RANGE} \otimes \text{fftw\_plan\_dft\_1d}_{N\_DOP}^{\text{inplace}}))$$

$$(I_{N\_CHAN} \otimes (I_{N\_RANGE} \otimes$$
$$(L_{N\_RANGE}^{N\_RANGE \cdot N\_DOP} \circ fftw\_plan\_dft\_1d_{N\_DOP})))$$

（a）领域语言描述图5（a）代码　　　　（b）领域语言描述图5（b）代码

图6：对使用领域语言描述的快速傅里叶变换进行优化[72]

## 6. 总结与展望

为作为高性能计算的一个重要目标，性能可移植的概念已经出现了很多年。为了突破摩尔定律的限制，当前超算系统的计算节点通常将通用处理器与一个或多个协加速器在主板或片上相连，通用处理器控制调度计算任务，而协加速器执行并行计算任务。当前超级计算机大多采用通用GPU以及Intel Xeon Phi作为协加速器，近年来，随着深度学习领域的不断发展，专用的神经网络加速芯片也开始不断涌现。不同架构的协加速器的出现及在超级计算机中的广泛应用，对程序在不同计算设备之间的性能可移植性提出了挑战。

本文从编程模型、串行代码自动并行化、并行代码自动转换等方面对当前的性能可移植技术进行了归纳介绍，同时在文章的最后还总结了如何通过合理使用科学计算库函数，达到提升程序的性能以及性能可移植性的目标。对于能够满足性能可移植要求的编程模型，抽象层次低的编程模型可以帮助有经验的开发人员实现高效的代码，但是也导致了其使用难度较高，而抽象层次高的编程模型降低了代码的开发难度，但对编译器提出了较高的要求。由于技术的限制，串行代码自动并行化技术目前只能在一定的范围内被使用。由于并行代码自动转换技术在理论和技术上不存在很高的难度，当前已经有较多研究结合目标计算设备的架构特点重新实现并行代码中已有的并行策略以及存储优化方法。综上所述，不同应用场景对应着不同的程序性能可移植的实现技术，程序开发人员为自己的程序选择性能可移植方案，其实是在各种约束条件下对编程效率以及优化效果进行平衡取舍。

根据以往研究成果和目前正在进行的研究，可以得出以下结论：1）程序性能可移植技术涵盖范围非常广泛，每一种技术都有其最适用的范围。程序开发人员为自己的程序选择性能可移植方案，其实是在各种约束条件下对编程效率以及优化效果进行平衡取舍；2）将算法描述以及优化解耦和对于实现性能可移植有着重要意义，当前仅有少数领域编程语言能够将算法描述以及优化解耦，并提供让编程人员介入代码优化的接口。如何通过设计新的编程模型，在更广的范围内实现高性能计算程序算法描述以及优化的解耦，还需要进行重点研究；3）串行代码自动并行化是实现代码性能可移植最直接的方法，其中多面体编译技术已经可以将符合条件的串行代码自动转换成目标计算设备上的并行代码，遗传编程对串行代码的限制则小的多，但是其产生的代码的可解释性也差许多。如何将自动并行化技术集成到编程模型中，提高程序的开发效率还需要更进一步的研究。

# References:

[1]. 什么才能促成性能的移植？ http://www.gpuworld.cn/article/show/538.html, 2016.
[2]. Dagum, L. and R. Menon, OpenMP: an industry standard API for shared-memory programming. IEEE Computational Science & Engineering, 1998. 5(1): p. 46-55.
[3]. Forum, M.P., MPI: A Message-Passing Interface Standard. 1994: University of Tennessee.
[4]. Blumofe, R.D., et al., Cilk: An Efficient Multithreaded Runtime System. Journal of Parallel & Distributed Computing, 1996. 37(1): p. 55-69.
[5]. TOP500.
[6]. Munshi, A. The OpenCL specification. in 2009 IEEE Hot Chips 21 Symposium (HCS). 2009.
[7]. Steuwer, M., P. Kegel and S. Gorlatch. SkelCL - A Portable Skeleton Library for High-Level GPU Programming. in ieee international symposium on parallel & distributed processing, workshops and phd forum. 2011.
[8]. Szuppe, J. Boost.Compute: A parallel computing library for C++ based on OpenCL. in International Workshop on Opencl. 2016.
[9]. Bolt C++ Template Library.
[10]. Potter, R., et al., Kernel composition in SYCL. 2015.
[11]. oneAPI.
[12]. Edwards, H.C., C.R. Trott and D. Sunderland, Kokkos: Enabling manycore performance portability through polymorphic memory access patterns. Journal of Parallel & Distributed Computing, 2014. 74(12): p. 3202-3216.
[13]. Hornung, et al., The RAJA Portability Layer: Overview and Status. 2014.
[14]. Mudalige, G.R., et al. OP2: An active library framework for solving unstructured mesh-based applications on multi-core and many-core architectures. in Innovative Parallel Computing. 2012.
[15]. Giles, M.B., et al., Designing OP2 for GPU architectures. Journal of Parallel & Distributed Computing, 2013. 73(11): p. 1451-1460.
[16]. Wienke, S., et al. OpenACC - First Experiences with Real-World Applications. in Proceedings of the 18th international conference on Parallel Processing. 2012.
[17]. Martineau, M., et al. Pragmatic Performance Portability with OpenMP 4.x. in International Workshop on Openmp. 2016.
[18]. Dolbeau, R., HMPPTM: A hybrid multi-core parallel programming environment., in General Purpose Processing on Graphics Processing Units. 2007.
[19]. Unat, D., C. Xing and S.B. Baden. Mint: Realizing CUDA performance in 3D stencil methods with annotated C. in International Conference on Supercomputing. 2011.
[20]. Chamberlain, B.L., D. Callahan and H.P. Zima, PARALLEL PROGRAMMABILITY AND THE CHAPEL LANGUAGE. International Journal of High Performance Computing Applications, 2007. 21(3): p. 291-312.
[21]. Sidelnik, A., et al. Performance Portability with the Chapel Language. in Parallel & Distributed Processing Symposium (IPDPS), 2012 IEEE 26th International. 2012.
[22]. Charles, P., et al. X10: An object-oriented approach to Non-Uniform Cluster Computing. in Proceedings of the 20th Annual ACM SIGPLAN Conference on Object-Oriented Programming, Systems, Languages, and Applications, OOPSLA 2005, October 16-20, 2005, San Diego, CA, USA. 2005.
[23]. Chen, T., et al., TVM: An Automated End-to-End Optimizing Compiler for Deep Learning.
[24]. XLA: Optimizing Compiler for Machine Learning. https://www.tensorflow.org/xla. 2017.

[25]. John, N., Scalable Parallel Programming with CUDA. Queue, 2008. 6(2): p. 1-9.
[26]. 伍明川等, 面向神威·太湖之光的国产异构众核处理器OpenCL编译系统. 计算机学报, 2018. 41(10): 第64-78页.
[27]. Rahman, R., OpenCL on Xeon Phi. Intel® Xeon Phi™ Coprocessor Architecture & Tools, 2013.
[28]. Lamb, C. OpenCL for NVIDIA GPUs. in IEEE Hot Chips 21 Symposium. 2009.
[29]. Berryhill, A., AMD Updates OpenCL SDK. Computer Reseller News, 2011(1312): p. p.52.
[30]. Gregory, K. and A. Miller, C++ AMP : accelerated massive parallelism with Microsoft Visual C++. Microsoft Press Corp, 2012. 37(5): p. 475-482.
[31]. HEROUX, M.A., et al., An Overview of the Trilinos Project. Acm Transactions on Mathematical Software, 2005. 31(3): p. 397-423.
[32]. Burgess, D.A., P.I. Crumpton and M.B. Giles. A Parallel Framework for Unstructured Grid Solvers. in Proceedings of the Second European Computational Fluid Dynamics Conference (1994). 1994.
[33]. Crumpton, P. and M. Giles, Multigrid aircraft computations using the OPlus parallel library. Parallel Computational Fluid Dynamics, 1996: p. 339-346.
[34]. Karypis, G., K. Schloegel and V. Kumar. Parmetis: Parallel Graph Partitioning and Sparse Matrix Ordering Library. 2003.
[35]. Chevalier, C. and F. Pellegrini. PT-Scotch: A tool for efficient parallel graph ordering. in parallel computing. 2008.
[36]. Balogh, G.D., et al., OP2-Clang: A Source-to-Source Translator Using Clang/LLVM LibTooling, in 2018 IEEE/ACM 5th Workshop on the LLVM Compiler Infrastructure in HPC (LLVM-HPC). 2018.
[37]. Jacobsen, N., LLVM supported source-to-source translation Translation from annotated C/C++ to CUDA C/C++. 2016.
[38]. Cunningham, D., R. Bordawekar and V. Saraswat, GPU programming in a high level language: compiling X10 to CUDA, in ACM SIGPLAN. 2011.
[39]. Chamberlain, B.L., D. Callahan and H.P. Zima, PARALLEL PROGRAMMABILITY AND THE CHAPEL LANGUAGE. International Journal of High Performance Computing Applications, 2007. 21(3): p. 291-312.
[40]. B. L. Chamberlain, et al., Authoring User-Defined Domain Maps in Chapel, in Cray Users Group Conference (CUG). 2011.
[41]. B. L. Chamberlain, S. J. Deitz and D. Iten, User-defined Data Distributions in Chapel: Philosophy and Framework., in Workshop on Hot Topics in Parallelism. 2010.
[42]. Sidelnik, A., et al. Performance Portability with the Chapel Language. in international parallel and distributed processing symposium. 2012.
[43]. Bolten, M., et al., Algebraic description and automatic generation of multigrid methods in SPIRAL. Concurrency and Computation: Practice and Experience, 2017. 29(17).
[44]. Ragan-Kelley, J.M., et al. Halide: A Language and Compiler for Optimizing Parallelism, Locality, and Recomputation in Image Processing Pipelines. in Proceedings of the 34th ACM SIGPLAN conference on Programming language design and implementation. 2013.
[45]. 赵捷, 李颖颖与赵荣彩, 基于多面体模型的编译"黑魔法". 软件学报, 2018. Vol.29Issue(8): 第2371-2396页.
[46]. Pop, S., et al. GRAPHITE: Polyhedral analyses and optimizations for GCC. in GCC Developper's Summit. 2006.
[47]. Trifunovic, K., et al., GRAPHITE Two Years After: First Lessons Learned From Real-World Polyhedral Compilation. Gcc Research Opportunities Workshop, 2010.
[48]. Grosser, T., A. GROESSLINGER and C. LENGAUER, POLLY — PERFORMING POLYHEDRAL OPTIMIZATIONS ON A LOW-LEVEL INTERMEDIATE REPRESENTATION. Parallel Processing Letters, 2013. 22(04): p. 28.
[49]. Bondhugula, U., et al. A practical automatic polyhedral parallelizer and locality optimizer. in programming language design and implementation. 2008.
[50]. Konstantinidis, A., et al. Parametric GPU Code Generation for Affine Loop Programs. in languages and compilers for parallel computing. 2013.
[51]. Verdoolaege, S., et al. Polyhedral parallel code generation for CUDA. in high performance embedded architectures and compilers. 2013.
[52]. Verdoolaege, S., PPCG and Pencil Compiler Design. 2016.
[53]. Grosser, T. and T. Hoefler. Polly-ACC Transparent compilation to heterogeneous hardware. in International Conference on Supercomputing. 2016.

[54]. Vasilache, N., et al., Tensor Comprehensions: Framework-Agnostic High-Performance Machine Learning Abstractions. 2018.
[55]. Koza, J.R., Genetic programming as a means for programming computers by natural selection. Statistics & Computing, 1994. 4(2): p. 87-112.
[56]. Koza and JohnR, Genetic programming : on the programming of computers by means of natural selection. 1992: MIT Press.
[57]. Ryan, C., J.J. Collins and M. O'Neill. Grammatical Evolution: Evolving Programs for an Arbitrary Language. in Genetic Programming, First European Workshop, EuroGP'98, Paris, France, April 14-15, 1998, Proceedings. 1998.
[58]. Langdon, W.B. and M. Harman. Evolving a CUDA Kernel from an nVidia Template. in Evolutionary Computation (CEC), 2010 IEEE Congress on. 2010.
[59]. Gopinath, C., R.M.A. Azad and C. Ryan. Multi-core GE : Automatic Evolution of CPU Based Multi-core Parallel Programs. in Proceedings of the 2014 conference companion on Genetic and evolutionary computation companion. 2014.
[60]. Chennupati, G., R.M.A. Azad and C. Ryan, Automatic Evolution of Parallel Sorting Programs on Multi-cores. 2015.
[61]. Gopinath, C., R.M.A. Azad and C. Ryan. Performance Optimization of Multi-Core Grammatical Evolution Generated Parallel Recursive Programs. in Genetic & Evolutionary Computation Conference. 2015.
[62]. Gopinath, C., Grammatical evolution + multi-cores = automatic parallel programming!. 2015.
[63]. Programming, G., Grammatical Evolution. Evolutionary Computation IEEE Transactions on, 2001. 5(4): p. 349-358.
[64]. Nawata, T. and R. Suda. APTCC : Auto parallelizing translator from C to CUDA. in international conference on conceptual structures. 2011.
[65]. 李雁冰等, 一种面向异构众核处理器的并行编译框架. 软件学报, 2019. Vol.30Issue(4): 第981-1001页.
[66]. Lee, S. and R. Eigenmann. OpenMPC: Extended OpenMP Programming and Tuning for GPUs. in High Performance Computing, Networking, Storage & Analysis. 2010.
[67]. Lee, S., S.J. Min and R. Eigenmann, OpenMP to GPGPU: a compiler framework for automatic translation and optimization. Acm Sigplan Notices, 2008. 44(4): p. 101-110.
[68]. 王燕燕, OpenMP-to-OpenCL代码自动转换工具的设计与实现, 2015, 吉林大学.
[69]. Stratton, J.A., S.S. Stone and W.W. Hwu. MCUDA: An Efficient Implementation of CUDA Kernels for Multi-core CPUs. in languages and compilers for parallel computing. 2008.
[70]. Denny, J.E., S. Lee and J.S. Vetter, Clacc: Translating OpenACC to OpenMP in Clang, in 5th Workshop on the LLVM Compiler Infrastructure in HPC. 2016.
[71]. Sultana, N., et al. From OpenACC to OpenMP 4: Toward Automatic Translation. in Xsede16 Conference on Diversity. 2016.
[72]. 李韦等, 提升高性能计算程序性能可移植性的领域特定语言. 高技术通讯, 2020(2): 第141-149页.
[73]. Franz Franchetti, et al., SPIRAL: Extreme Performance Portability, in IEEE special issue on From High Level Specification to High Performance Code. 2018.
[74]. Linton, S., GAP: groups, algorithms, programming. Acm Communications in Computer Algebra, 2007. 41(3): p. 108-109.